# Study of similarity rules for electromagnetic process in partially ionized plasmas

Jiansheng Yao[1*], Yingkui Zhao[1*], Hantian Zhang[1], Difa Ye[1], Biyao Ouyang[1]

[1]Institute of Applied Physics and Computational Mathematics, Beijing 100088, China

**Correspondence to:** Jiansheng Yao; Yingkui Zhao

**Email**: ab135794@mail.ustc.edu.cn; zhao_yingkui@iapcm.ac.cn

**Abstract**

As proved by previous study, the similarity of electromagnetic processes in plasmas will be violated by Coulomb collisions between electron and ions. Therefore, there is no similarity in highly ionized collisional plasma. But the situation will be completely different for collisional plasmas with low ionization degree. The main collision type will change from electron-ion Coulomb collision to electron-molecule collision, and a new variable (the number density of neutral molecules) will be introduced into the similarity constraint, which can increase the degree of freedom. Thus, in this condition, the similarity restriction caused by the collision process doesn't conflict with the other restrictions. Therefore, the similarity for electromagnetic process in collisional plasmas can be valid for partially ionized plasmas. In this paper, we propose the similarity in partially ionized plasmas, and prove it via particle in cell/Monte Carlo (PIC/MCC) simulation. Our research has a wide range of engineering applications.

# 1.Introduction:

As we all know, the similarity has been widely used to simulate large-scale physical phenomena with much smaller scale models [1-3]. What's more, the similarity can also be applied to compare different experiments on the same physical phenomenon. In plasma physics, the similarity has been an attractive topic, for example: in fusion experiments [4-5], large-scale devices have exorbitant costs, so it is critical to use similarity laws for early design on small-scale devices; Besides, the similarity enable us to simulate large-scale astrophysical plasma phenomena at the laboratory [6-7].

Using $\pi$-theorem [8], A. Beiser and B. Raab [9] first proposed hydrodynamic similarity rules in plasma physics. After that, Lacina [10] extensively investigated similarity rules in plasma physics. In his study, particles' motion is described by the Boltzmann equation with a binary collision term. According to his theory, six invariants couldn't be satisfied at the same time, thus there is no similarity among experiments with collisional electromagnetic process. To be specific, the invariant corresponding to the collision contradicts to other invariants corresponding to the evolution of particles and fields. Therefore, similarity rules have been only applied in experiments with collisional electrostatic process [11-14] or collisionless electromagnetic process [15-17]. Unfortunately, these pioneering works on the similarity of plasmas appear to have gone unnoticed. A recent review[18] on the similarity of magnetic fusion experiments highlighted the significance of Lacina.'s work.

Different from the fully ionized plasmas studied by previous researchers, it is frequently necessary in engineering to study the similarity in partially ionized plasmas. The difference in the degree of ionization makes the types of collisions that need to be considered in thees two situations completely different. In fully or highly ionized plasmas, the collision type is mainly electron-ion Coulomb collision, while in low ionized plasmas, the collision type is mainly electron-molecule collision. The different types of collisions will produce different similarity invariants, thus the similarity in partially ionized plasmas must be reconsidered. The system electromagnetic pulse[19-22] (SGEMP), a classic collisional electromagnetic process in low ionized plasmas, will be used as a research topic in this paper to investigate the similarity in the collisional electromagnetic process of partially ionized plasmas.

The System generated electromagnetic pulse (SGEMP) is the transient electromagnetic field produced by energetic electrons emitted from the surfaces of metals. The main physical processes involved in the SGEMP are: X/$\gamma$ rays produce high-energy electrons (are also called primary electrons) from the metal surfaces via the photoionization/Compton effects; the motion of these electrons produces the transient electromagnetic field. This process can be described using a particle in cell (PIC) simulation model [23-24]. The SGEMP is more sophisticated when the electronic systems operate in a gaseous environment [25-26], the ionization collision between energetic electrons and gas molecules would produce secondary electrons. Forced by electric/magnetic fields, these electrons would move and constitute secondary electrons and weaken electric/magnetic fields. This process is generally described with Monte Carlo simulation model [27-29]. Different from electromagnetic field propagated from outside of the system[30], the SGEMP is hard to mitigated via Faraday cage. Due to its destruction of electronic devices and power infrastructure, the SGEMP has been an important research area of electromagnetic protection since 1960s.

There have been studies [31-32] on the similarity of SGEMP in a vacuum, and obtained the similarity consistent with Lacina's theory, but there are no similarity of SGEMP in gaseous environment yet. In this paper, as a classic collisional electromagnetic process in partially ionized plasmas, the similarity of SGEMP in gaseous environment will be investigated to prove our similarity rules. The paper is organized by following sections: the discussion of similarity rules is presented in section 2; the description of the simulation model is presented in section 3; the simulation results are illustrated in section 4; summary and discussion are presented in section 5.

## 2. Similarity rules

According to Buckingham $\pi$ theorem [8], physically equations can be recast in terms of dimensionless combinations of the physical variables, if these invariants remain unchanged, the solution of equations will also remain unchanged. This serves as a valuable guideline for experiments. For example, make the Reynolds number remain constant, we only need a small aircraft model to test the aircraft's aerodynamic performance in the wind tunnel. We will derive the similarity in this section based on the theory of Ryutov and Remington[17] using the international system of units due to the differences in collision types and units system with previous study[10].

The motion of particles is described via Fokker-Planck equations,

$$\frac{\partial f_s}{\partial t} + \vec{v} \cdot \frac{\partial f_s}{\partial \vec{r}} + \frac{q_s}{m_s}(\vec{E} + \vec{v} \times \vec{B}) \cdot \frac{\partial f_s}{\partial \vec{v}} = (\frac{\partial f_s}{\partial t})_{coll}, \tag{1}$$

in which, the term $(\partial f_s / \partial t)_{coll}$ represents the collision between species s and k, and can be written as

$$\begin{aligned}(\frac{\partial f_s}{\partial t})_{coll} &= \iiint f_s(\vec{v}_1) f_k(\vec{v}_2) |\vec{v}_1 - \vec{v}_2| \sigma_{sk}(\vec{v}_1, \vec{v}_2; \vec{v}_3, \vec{v}) d^3\vec{v}_1 d^3\vec{v}_2 d^3\vec{v}_3 \\ &- \iiint f_s(\vec{v}) f_k(\vec{v}_3) |\vec{v} - \vec{v}_3| \sigma_{sk}(\vec{v}, \vec{v}_3; \vec{v}_1, \vec{v}_2) d^3\vec{v}_1 d^3\vec{v}_2 d^3\vec{v}_3. \end{aligned} \tag{2}$$

The first term on the right denotes an increase in the distribution of s particles in velocity space $\vec{v}$, $\sigma_{sk}(\vec{v}_1, \vec{v}_2; \vec{v}_3, \vec{v})$ is the cross section between species s and k, the initial velocities of species s and k are $\vec{v}_1$ and $\vec{v}_2$, after collision, the final states of species s and k are $\vec{v}$ and $\vec{v}_3$, respectively. The second term on the right represents the decrease in the distribution of s particles in velocity space $\vec{v}$, the initial velocities of species s and k are $\vec{v}$ and $\vec{v}_3$, the final states of species s and k are $\vec{v}_1$ and $\vec{v}_2$, respectively.

The self-consistent electric field $\vec{E}$ and magnetic field $\vec{B}$ in Eq.(A1) are given by the Maxweillian eqations:

$$\frac{\partial \vec{B}}{\partial t} = -\nabla \times \vec{E}, \tag{3}$$

$$\varepsilon_0 \frac{\partial \vec{E}}{\partial t} = \frac{1}{\mu_0} \nabla \times \vec{B} - \vec{j}, \tag{4}$$

$$\nabla \cdot \vec{E} = \frac{\rho}{\varepsilon_0}, \tag{5}$$

where current density is $\vec{j} = \sum_s e_s \int \vec{v} f_s d^3\vec{v}$, charge density is $\rho = \sum_s e_s \int f_s d^3\vec{v}$. The physical quantity $M$ is composed of the product of its dimension $M_0$ and a dimensionless variable $\hat{M}$, i.e. $M = M_0 \hat{M}$. Therefore, Eq.(1) can be rewritten as follows:

$$\frac{\partial \hat{f}_s}{\partial \hat{t}}(\frac{f_{s0}}{t_0}) + \hat{\vec{v}} \cdot \frac{\partial \hat{f}_s}{\partial \hat{\vec{r}}}(\frac{\vec{v}_0 f_{s0}}{\vec{r}_0}) + (\hat{\vec{E}} + \hat{\vec{v}} \times \hat{\vec{B}}) \cdot \frac{\partial \hat{f}_s}{\partial \hat{\vec{v}}}[\frac{q_s}{m_s}(E_0 + \vec{v}_0 \times \vec{B}_0)\frac{f_{s0}}{\vec{v}_0}] = \\ (\frac{\partial \hat{f}_s}{\partial \hat{t}})_{coll} \cdot F(f_{s0}, f_{k0}, v_0, \sigma_{sk}). \tag{6}$$

Eq.(6) is deformed to get

$$\frac{\partial \hat{f}_s}{\partial \hat{t}} + \hat{\vec{v}} \cdot \frac{\partial \hat{f}_s}{\partial \hat{\vec{r}}}(\frac{v_0 t_0}{r_0}) + (\hat{\vec{E}} + \hat{\vec{v}} \times \hat{\vec{B}}) \cdot \frac{\partial \hat{f}_s}{\partial \hat{\vec{v}}}[\frac{q_s}{m_s}(E_0 + v_0 B_0)\frac{t_0}{v_0}] = \\ (\frac{\partial \hat{f}_s}{\partial \hat{t}})_{coll} \cdot F(f_{s0}, f_{k0}, v_0, \sigma_{sk})\frac{t_0}{f_{s0}}, \tag{7}$$

where $F(f_{s0}, f_{k0}, v_0, \sigma_{sk})\frac{t_0}{f_{s0}}$ denotes the collision of species s and k, and has the following formation:

$$F(f_{s0}, f_{k0}, v_0, \sigma_{sk})\frac{t_0}{f_{s0}} = f_{k0} t_0 v_0^4 \Sigma_{sk}, \tag{8}$$

where $\Sigma_{sk}$ is the total cross section $\Sigma_{sk} = \iint \sigma_{sk}(\vec{v}_1, \vec{v}_2; \vec{v}_3, \vec{v}_4) d^3\vec{v}_1 d^3\vec{v}_2$.

According to Eqs. (6) and (7), the particle trajectory is similar if the following dimensionless physical quantities remain unchanged.

$$C_1 = \frac{v_0 t_0}{r_0}, \quad C_2 = \frac{q_s E_0 t_0}{m_s v_0}, \quad C_3 = \frac{q_s B_0 t_0}{m_s}, \quad C_4 = f_{k0} t_0 v_0^4 \Sigma_{sk}., \tag{9}$$

Similarly, Eqs. (3) and (4) can be rewritten as follows:

$$\frac{\partial \hat{\vec{B}}}{\partial \hat{t}} \frac{B_0}{t_0} = -\hat{\nabla} \times \hat{\vec{E}} \frac{E_0}{r_0}, \tag{10}$$

$$\frac{\partial \hat{\vec{E}}}{\partial \hat{t}} (\varepsilon_0 \frac{E_0}{t_0}) = \hat{\nabla} \times \hat{\vec{B}} (\frac{B_0}{\mu_0 r_0}) - \sum_s \hat{\vec{v}} \hat{f}_s d^3 \hat{\vec{v}} (e_s f_{s0} v_0^4), \tag{11}$$

$$\hat{\nabla} \cdot \hat{\vec{E}} (\frac{E_0}{r_0}) = \int \hat{f}_s d^3 \hat{\vec{v}} (\frac{1}{\varepsilon_0} e_s f_{s0} v_0^3). \tag{12}$$

Eqs. (9) and (10) illustrate that the field are similar if the following three quantities remain constant:

$$C_5 = \frac{B_0 t_0}{E_0 r_0 \varepsilon_0 \mu_0}, \quad C_6 = \frac{e_s f_{s0} v_0^4 t_0}{\varepsilon_0 E_0}, \quad C_7 = \frac{f_{s0} v_0^3 e_s r_0}{\varepsilon_0 E_0}, \quad C_8 = \frac{E_0 t_0}{B_0 r_0}. \tag{13}$$

Substituting $C_1$ into $C_2 - C_4$, we get

$$C_2' = \frac{q_s E_0 r_0}{m_s v_0^2}, \quad C_3' = \frac{q_s B_0 r_0}{m_s v_0}, \quad C_4' = f_{k0} r_0 v_0^3 \Sigma_{sk}. \tag{14}$$

We can obtain $C_7$ and $C_8$ via combing other six invariants, thus $C_7$ and $C_8$ are not independent invariants and can be neglected. Substituting $C_1$, $C_2' - C_4'$ into $C_5 - C_6$ and eliminating $t_0$ and $r_0$, we will obtain

$$C_5' = \frac{c^2}{v_0^2}, \quad C_6' = \frac{f_{s0} v_0^5 m_s}{\varepsilon_0 E_0^2} \tag{A15}$$

Invariants $C_1$ and $C_2' - C_6'$ differ slightly with the invariants given by Lacina (1971), this difference caused by different system of units applied in their study.

In general, the incident particles are the same for different experiments, thus, $m_s$ and $q_s$ should be unchanged. In addition, the type of target particles also remains unchanged, this indicates the total cross section $\Sigma_{sk}$ also stay the same. The speed of light in the vacuum c is the same in different experiments, thus the velocity of particles in different experiments much keep the same due to the restriction of the invariant $C_5$, this restriction stems from electromagnetic equation Eq.4, which does not exist in the electrostatic simulation. Assuming that the two experimental systems' scales satisfy $L_1 = \xi L_0$, we obtain $t_1 = \xi t_0$ from $C_1$ considering that $v_1 = v_0$. Thus, derived from the constriction of invariants $C_2^{'}$ and $C_3^{'}$, electric field and magnetic field in these two experiments should satisfy $E_1 = E_0 \xi^{-1}$ and $B_1 = B_0 \xi^{-1}$. According to the invariant $C_6^{'}$, the number density of incident particles $n_s = f_{s0} v_0^3$ satisfy $n_{s1} = n_{s0} \xi^{-2}$.

The most interesting is the similarity of the collision process, which is related to the invariant $C_4^{'}$, the number density of target particles $n_k = f_{k0} v_0^3$ should satisfy $n_{k1} = n_{k0} \xi^{-1}$. This constriction couldn't be satisfied for fully (or highly) ionized plasmas, in which, the main collision type is coulomb collision, the target particle is ion, thus $n_k = n_s$ due to the requirement of electrical neutrality. Obviously, the similarity derived from the invariant $C_4^{'}$ ($n_{s1} = n_{s0} \xi^{-1}$) is contradict to that from $C_6^{'}$ ($n_{s1} = n_{s0} \xi^{-2}$), thus, just as asserted by Lacina (1971), there is no similarity in this condition. However, the situation is different for plasmas with low ionization degree, in this condition, the main collision type will change to be the collision between electrons and molecules, the target particle is gas molecule. Therefore, a new parameter, the number density of target molecules $n_k$, is introduced into the system and capable of resolving the conflict between the invariants $C_4^{'}$ and $C_6^{'}$, and target molecules in these two experiments should satisfy $n_{k1} = n_{k0} \xi^{-1}$.

In conclusion, there are six invariants needed to be satisfied for electromagnetic process in partially ionized plasmas.

$$\Pi_1 = \frac{v_0 t_0}{r_0}, \quad \Pi_2 = \frac{q_s E_0 r_0}{m_s v_0^2}, \quad \Pi_3 = \frac{q_s B_0 r_0}{m_s v_0}, \quad \Pi_4 = f_{k0} r_0 v_0^3 \Sigma_{sk}, \quad \Pi_5 = \frac{c^2}{v_0^2}, \quad \Pi_6 = \frac{f_{s0} v_0^5 m_s}{\varepsilon_0 E_0^2}$$

Not all of these six invariants are always needed to be considered, table 1 shows parameters to be considered in various situations.

**Table 1. Invariants in different process**

|  | Electrostatic process | Electromagnetic process |
| --- | --- | --- |
| Collisionless plasmas | $\Pi_1, \Pi_2, \Pi_3, \Pi_6$ | $\Pi_1, \Pi_2, \Pi_3, \Pi_5, \Pi_6$ |
| Collisional plasmas | $\Pi_1, \Pi_2, \Pi_3, \Pi_4, \Pi_6$ | $\Pi_1, \Pi_2, \Pi_3, \Pi_4, \Pi_5, \Pi_6$ |

The above six invariants must be satisfied simultaneously for the electromagnetic process in a partially ionized plasma, yielding the following relationship, and we will get the following relationship (Table 2) based on these six invariants.

**Table 2. Quantities before and after scaling**

| Quantities | Original | Scaled |
| --- | --- | --- |
| Length/Time | $L, t$ | $\xi L, \xi t$ |
| Particle Mass/Charge/Velocity | $m_s, q_s, v$ | $m_s, q_s, v$ |
| Particle Energy/Velocity | $\varepsilon, v$ | $\varepsilon, v$ |
| Cross Section | $\sigma$ | $\sigma$ |
| Particles Number Density | $n$ | $\xi^{-2} n$ |
| Current | $J$ | $\xi^{-2}J$ |
| Target Molecules Number Density | $n_g$ | $\xi^{-1}n_g$ |
| Collision Frequency | $v$ | $\xi^{-1}v$ |
| Electric/Magnetic Field | $E, B$ | $\xi^{-1}E, \xi^{-1}B$ |

Woods and Wenaas[31] proposed similarity laws for SGEMP in a vacuum environment, which is consistent with that shown in Table 1 and 2. However, there is no similarity laws for SGEMP in gaseous environment by now, therefore we will investigate it in the following sections via simulations. What's more, as a classic electromagnetic process in partially ionized plasmas, the SGEMP in gaseous environment will be an excellent object to prove the similarity proposed in this paper.

3. Simulation model

The schematic of SGEMP simulation model is presented in the Fig. 1. A cylindrical cavity made of aluminum has an inner diameter $r$ and an inner length L. The cylinder's axis is parallel to the z-axis. In the SGEMP process, X-rays enter the cylinder from the bottom and ionize the metal surface, producing photoelectrons (are also called primary electrons) after they are absorbed. In this study, we are only concerned with similarity rules in the electromagnetic process of partially ionized plasmas, rather than the detailed physical process of SGEMP. Therefore, the photoelectrons produced by ionization are assumed to be mono-energetic and move along the z-axis.

On each step, 200 macro electrons with energies of 10keV are emitted from the bottom of the cavity. Every macro particle represents $10^7\xi$ real particles, where $\xi$ is the scale factor. Therefore, according to table 2, the particle number density is $\xi^{-2}n$ $m^{-3}$ after scaling. After being injected into the cavity, electrons will collide with the air molecules. Collision types are: elastic collision between electrons and oxygen/nitrogen molecules, ionization between electrons and oxygen/nitrogen molecules, and adsorption between electrons and oxygen molecules. The energy-dependent cross sections obtained from LXcat (https://nl.lxcat.net/data/set_type.php) are presented in Fig. 2. The number density of nitrogen and oxygen molecules are $0.8 \times 10^{22} \xi^{-1} \ m^{-3}$ and $0.2 \times 10^{22} \xi^{-1} \ m^{-3}$, respectively.

The injected primary electrons ionize air molecules to produce a large number of secondary electrons. The increasing number of secondary electrons will cause the Debye length $\lambda_D = \sqrt{\varepsilon_0 k_B T_e / n_e e^2}$ ( $\varepsilon_0$ is the vacuum permittivity, $k_B$ is the Boltzmann constant) to decrease, destorying the simulated stable conditions $\Delta x < 3\lambda_D$ due to 'self-heating'. To avoid this, we need to select a smaller spatial step size. In this study, the maximum number density of secondary electrons produced by is about $10^{16} \xi^{-2} \ m^{-3}$, and the mean energy of secondary electrons is 10 $eV$, thus the Debye length in this condition is $\lambda_D \approx 2.3 \times 10^{-4} \xi^{-1} \ m$, and spatial size should satisfy $\Delta x < 7 \times 10^{-4} \xi^{-1} \ m$. It is difficult to solve 3D large-scale simulations with such a small spatial step size. Therefore, an energy conservation algorithm [31] is needed to extend the spatial length while maintaining the calculation's stability. This algorithm calculates current and charge density on the grid using second-order interpolation, and interpolates the field on the grid to the particle's position using first-order interpolation. According to previous study [34], this algorithm can ensure stable calculation under the condition of $\Delta x \sim 1000 \lambda_D$, and has been widely used in the simulation of SGEMP [35-36]. In our simulation, we also adopt this method. The grid number is $n_x \times n_y \times n_z = 32 \times 32 \times 54$, the size of simulation domain is $0.32 \xi \ m \times 0.32 \xi \ m \times 0.54 \xi \ m$, where $\xi$ is the scale factor. The time step is $1.14 \times 10^{-11} \xi \ s$.

In this study, to prove the similarity in the electromagnetic process of partially ionized plasmas proposed in section 2, we will investigate the similarity of SGEMP via comparing quantities at two different scaled factors $\xi = 1$ and $\xi = 0.1$. According to the Table 1 and 2, in these two experiment with different scaled factors, if the injected electrons and target gas satisfy the following relationship: injected electrons number density is $\xi^{-2}n_e$, velocity is $\xi^0 v_0$, and target gas number density $\xi^{-1}n_g$, then the electromagnetic field generated by electron movement would satisfy the following relations: electric field is $\xi^{-1}E$, magnetic field is $\xi^{-1}B$. We will investigate above relations in the next section.

4. Simulation Results

As shown in table 2, firstly, we need to ensure that the incident primary electrons meet similarity rules, that is: the speed of the electron remains unchanged, and the scaled number density of the electron becomes $\xi^{-2}n$ $m^{-3}$. Figures 3 (a) and (b) are spatial distribution of primary electrons at different scaled factors $\xi = 1$ and $\xi = 0.1$, respectively. And the corresponding time is $5.15 \times 10^{-10}$ s and $5.15 \times 10^{-9}$ s. Different colors represent different $v_z$. Obviously, in the cases of different scale factors, primary electrons have a similar spatial distribution and velocity. The electrons with higher velocity are distributed at the far end of the z axis, while the low velocity electrons are distributed near z=0. Fig. 3 (c) depicts the distribution of primary electrons' number density along the z direction at this moment. It should be noted that, the ordinate is $n_e \xi^{-2}$ $m^{-3}$ to facilitate the comparison of results. Obviously, the number density of primary electrons satisfies similarity rules. According to table 2, the scaled number density of target molecules should be $\xi^{-1}n_g$ to ensure collision process similarity. When aforementioned conditions are met, we will test whether the secondary electrons ionized by primary electrons follow similarity rules.

Figures 4 (a) and (b) are spatial distribution of secondary electrons at different scaled factors $\xi = 1$ and $\xi = 0.1$, respectively. And the corresponding time is $5.15 \times 10^{-10}$ s and $5.15 \times 10^{-9}$ s. Different colors represent different $v_z$. It is obvious that, in the cases of different scale factors, secondary electrons also have a similar spatial distribution and have the same velocity. Fig 4 (c) is the variation of secondary electrons' number density along the z direction at this moment. It is obvious that the number density of secondary electrons also satisfies similarity laws. What needs to be emphasized is that, in the calculation of secondary electrons, we use particle merging technology to speed up the program. In the merging process, electrons with similar speeds are combined into one macro particle, and the weight of the new macro particle is the sum of their respective weights. Therefore, when calculating changes in the number of secondary electrons, we must account for different weights of different particles.

In the following, we will investigate whether the electric/magnetic fields generated by the movement of electrons can satisfy similarity rules. Figure 5 (a)-(c) describe the spatial distribution of electric field $E_z$ at three moments when $\xi=0.1$. Figures 5 (d)-(f) correspond to Figures 5 (a)-(c), which depict the spatial distribution of the electric field $E_z$ at three different moments when $\xi=1$. It is worth noting that the time corresponding to Figure 5 (a)-(c) is 0.1 times that of Figure 5 (d)-(f). Obviously, the spatial distribution of the two rows of electric fields corresponds perfectly. It should be noted, however, that the colorbars in the two rows of electric fields are different. The colorbar range of the first row in Figure 5 is ten times that of the corresponding figure in the second row. This means that the electric field in small-scale is ten times larger than that in large-scale. We extract the electric field $E_z$ along the axis of the cylinder in Figure 5 to get Figure 6. In order to compare the two rows in Figure 5, we set the abscissa in Figure 6 as $z/\xi$ $m$, the ordinate as $E_z\xi$ $Vm^{-1}$, and the time as $t/\xi$ $s$. In the first stage of the simulation (Fig. 6(a)), the electric fields $E_z$ under the two different scale factors show a high degree of consistency. But, as shown in Figs. 6 (b) and (c), the electric fields $E_z$ at the left side violate the similarity, this is caused by the combination of secondary electrons, this can be suppressed via reducing combination frequency.

Figures 7 (a) and (b) are the evolution of the peak values of the electric field $E_z$ and the magnetic field $B_y$ in the analog domain, respectively. In the figure, the abscissa is $t/\xi$ $s$ and the ordinates are $E_z\xi$ $Vm^{-1}$ or $B_y\xi$ $T$. Obviously, both the electric field $E_z$ and the magnetic field $B_y$ satisfy similarity laws, but it should be noted that the similarity is not exactly satisfied at the late stage of the simulation, this phenomenon also emerges in the figure 6 (c), we will study it at the end of this section. We also test similarity rules at higher air pressure with $n_g = 4 \times 10^{22}\ m^{-3}$. Figures 8 (a) and (b) are the evolution of the peak values of the electric field $E_z$ and the magnetic field $B_y$ at higher air pressure. It is clear that similarity rules still hold at higher pressure. The peak values of electric and magnetic fields are $3.9 \times 10^5 \xi\ Vm^{-1}$ and $3.25 \times 10^{-4} \xi\ T$. Comparing Figures 7 and 8, it is found that the peak electric field at high pressure is smaller than that at low pressure. This is due to the higher density of secondary electrons produced by high-pressure ionization, which reduces the amplitude of the electric field [26].

The last remaining problem is the similarity is not exactly satisfied at the late stage of the simulation. Our investigation shows that this comes from the Monte Carlo sampling. If we remove the Monte Carlo process, the similarity will validate at all the time. The reason is that, a large number of electrons are absorbed by the boundary as the simulation progresses, the number of macro particles in the cavity reduces significantly at the late stage. At this time, the error caused by the process of the Monte Carlo sampling will rapidly magnify (it can be analogized to the decrease of the number of coin toss, the probability of heading up will deviate from 1/2). Beyond that, frequent combination of secondary electrons will also lead to the simulation errors.

To reduce the error caused by these two process, we increase the number of macro particles in the system and reduce the frequency of secondary electrons' merging. Specially, the number of injected macro particles increases 10 times (simultaneously, the number of real particles represented by a macro particle is reduced by 10 times to ensure that the system's total number of real particles remains constant). In addition, the frequency of merging secondary electrons is also greatly reduced, in previous model, secondary electrons begin to merge when the number in a cell exceeds 50, this value increase to 400 in the new simulation. As presented in Fig.9 (a), the similarity is violated after $t/\xi = 0.4$ s. This violation will be suppressed effectively in Fig.9 (b). This demonstrates that the error is caused by insufficient macro particles in the later stages of the simulation, and can be suppressed via increasing the number of macro particles in the system while decreasing the frequency of secondary electrons' combination. As shown in Fig.10, the violation of similarity at $t/\xi = 5.15e - 9\ s$ on the bottom ($z = 0$) of simulation domain can also be diminished via reducing the frequency of secondary electrons' combination.

In conclusion, for the electromagnetic process in collisional plasmas, keeping the incident particles and target particles unchanged, if incident particle density and the background gas density meet the similarity rules, the electric/magnetic fields generated by the particle motion also meet similarity rules.

## 5. Summary and Conclusion

As proved by previous study [10], there is no similarity of electromagnetic process of fully ionized collisional plasmas. However, as proved in this paper, the similarity is established for electromagnetic process of collisional plasmas with low ionization degree. This because the main collision type will change from electron-ion Coulomb collision in fully ionized plasmas to electron-molecule collision in partially ionized plasmas, and a new variable (the number density of neutral molecules) will be introduced into the similarity constraint, which can increase the degrees of freedom. Thus, in this condition, the similarity restriction caused by the collision process doesn't conflict with the other restrictions. In the absence of collision, the similarity in this paper is consistent with previous study [30]. Through investigating a classic electromagnetic process in collisional plasmas with low ionization degree- the SGEMP in gaseous environment, we prove the similarity in this paper.

In the simulation, we find that the similarity is not exactly satisfied at the late stage of the simulation. Further investigation reveals that the violation of the similarity is caused by the insufficient macro particles in the late stage. And this violation can be suppressed via increasing the number of macro particles in the system and decreasing the frequency of secondary electrons' combination.

Our research can be applied directly to engineering experiments. For example, when studying the SGEMP in aircraft cabins, according to similarity laws proposed in this paper, it can be reduced to the laboratory scale. But, according to our current knowledge, there will be three major factors in the experiment that undermine the similarity. The first factor is coulomb collision, in our study, the maximum ionization degree of injected electrons is $10^{-7}$ (the maximum ionization degree of secondary electrons is slightly higher and can reach $10^{-6}$), thus coulomb collision is very weak relative to the collision within electrons and molecules. But for conditions with higher ionization degree, according to previous analysis, the effect of coulomb collision increase and violate the similarity. The second factor violating the similarity may come from the interaction of electrons with cavity boundaries. In our simulation, particles are absorbed when they reach the boundary; however, in the experiment, electrons can be reflected by the boundary; furthermore, high-energy electrons can ionize the boundary, ejecting secondary electrons. The last factor is three-body interaction, only two-body collision processes are included in our theory and simulation, but three-body interactions in experiments must violate the similarity. These complicated process may also violate the similarity. We will study the influence of these factors in our future work.

**Acknowledgments:**

...This work is supported by the National Science Foundation of China 11822401, 12174034. Data underlying the results presented in this paper are not publicly available at this time but can be obtained from the authors upon reasonable request.

**Figure Captions:**

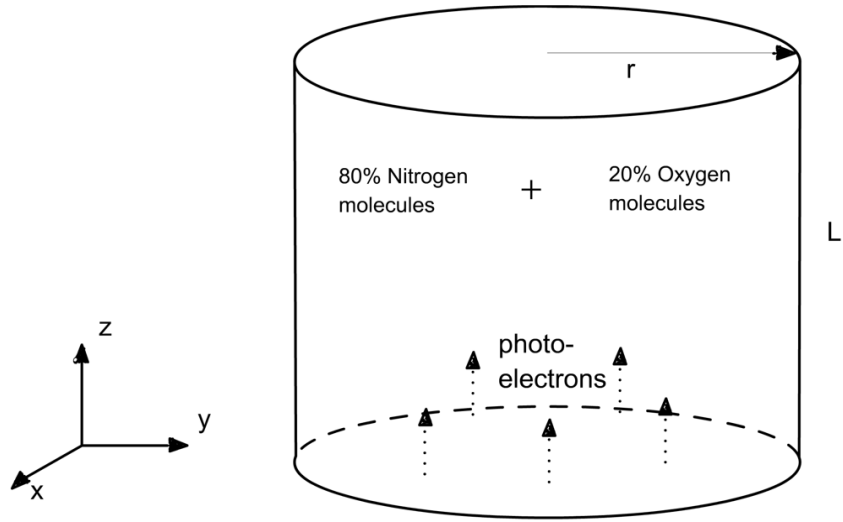

**Figure 1.** Schematic of the simulation model. The cylindrical surface is made of aluminum, electrons are emitted from the bottom into the cavity, and the cavity is filled with air (80% $N_2$+20% $O_2$), The axis of the cylinder is along the z direction.

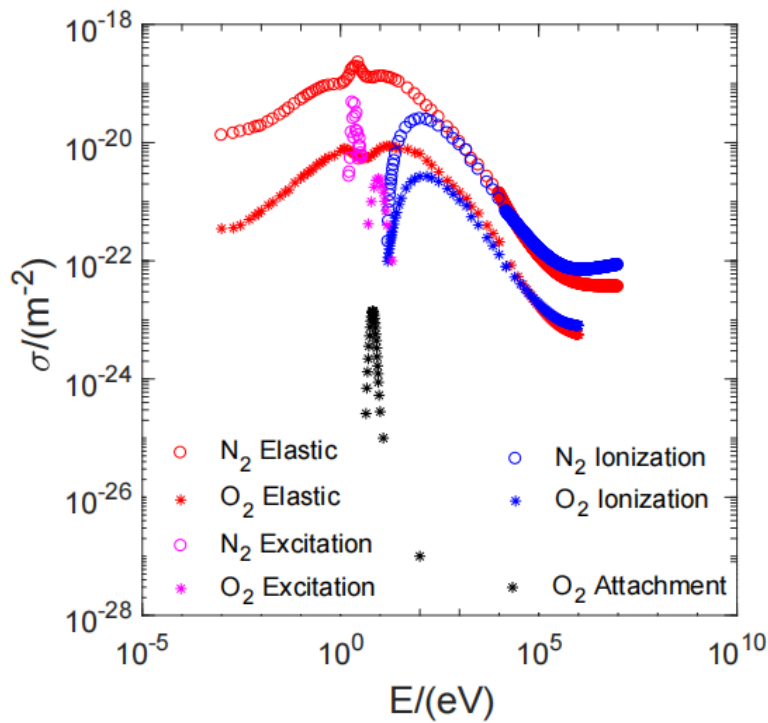

**Figure 2.** Cross section of electron-neutral collision. Collisions between electrons and $N_2/O_2$ molecules are represented by the dashed and solid lines, respectively. The red, blue, and black lines represent the elastic collision, ionization, and adsorption processes, respectively.

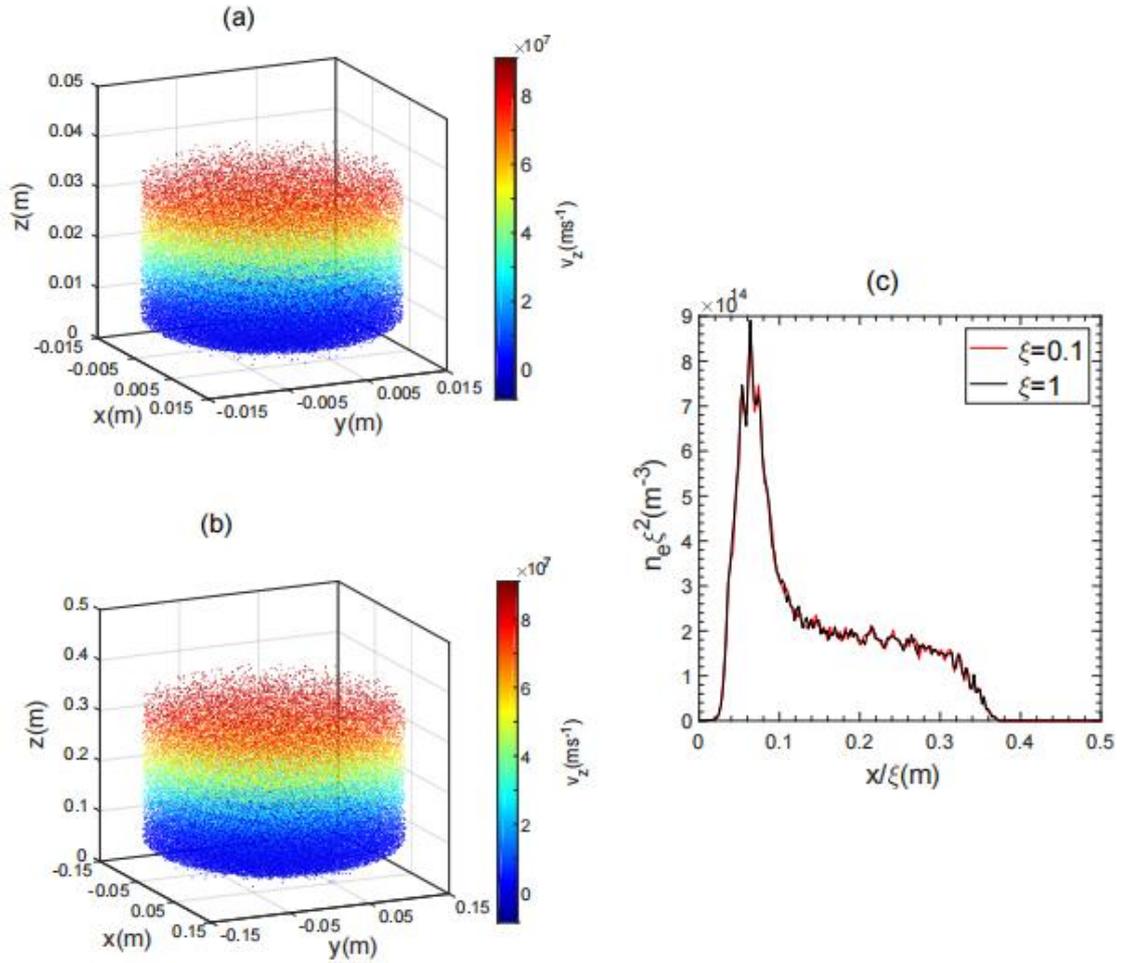

**Figure 3.** (a) the spatial distribution of primary electrons at $t = 5.15 \times 10^{-10}$ s for scale factor $\xi = 0.1$.

(b) the spatial distribution of primary electrons at $t = 5.15 \times 10^{-9} s$ for scale factor $\xi = 1$.

(c) the number density of primary electrons along z-axis at $t/\xi = 5.15 \times 10^{-9} s$. Different colors in (a) and (b) represent different $v_z$.

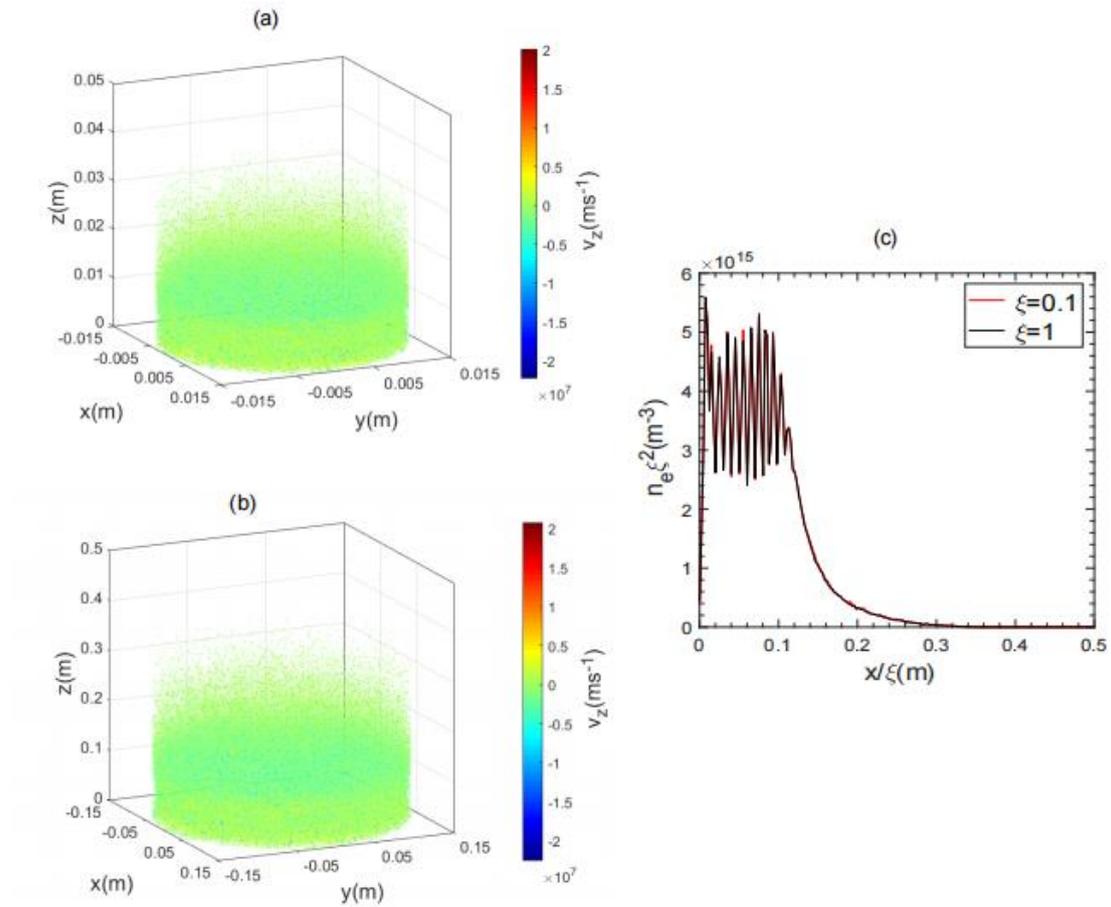

**Figure 4.** (a) the spatial distribution of secondary electrons at $t = 5.15 \times 10^{-10}$ s for scale factor $\xi = 0.1$.

(b) the spatial distribution of secondary electrons at $t = 5.15 \times 10^{-9} s$ for scale factor $\xi = 1$.

(c) the number density of secondary electrons along z-axis at $t/\xi = 5.15 \times 10^{-9} s$. Different colors in (a) and (b) represent different $v_z$.

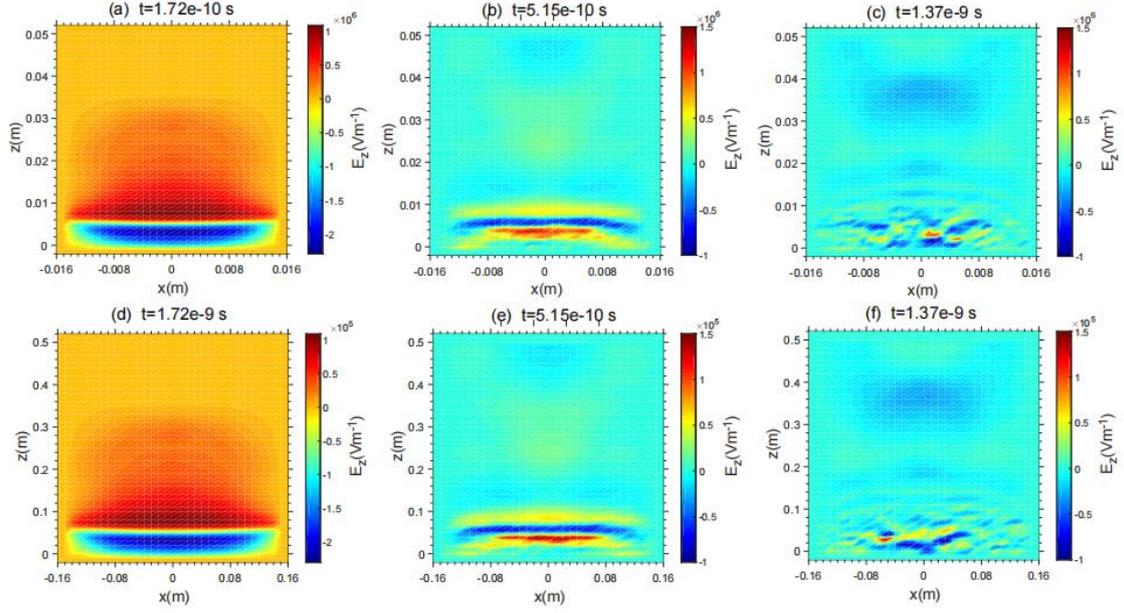

**Figure 5.** (a)-(c) describe the spatial distribution of electric field $E_z$ for scale factor $\xi = 0.1$ at three moments $t = 1.72 \times 10^{-10}$ s, $t = 5.15 \times 10^{-10}$ s and $t = 1.37 \times 10^{-9}$ s, respectively. (d)-(f) describe the spatial distribution of electric field $E_z$ for scale factor $\xi = 1$ at three moments $t = 1.72 \times 10^{-9}$ s, $t = 5.15 \times 10^{-9}$ s and $t = 1.37 \times 10^{-8}$ s, respectively.

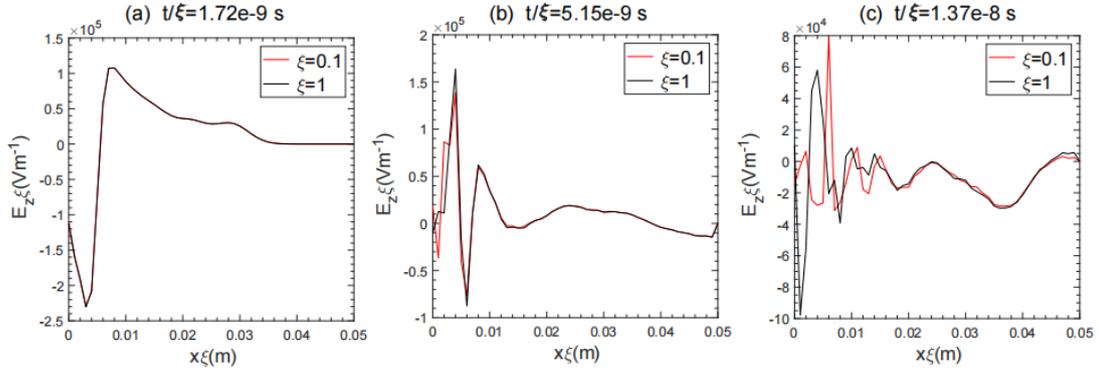

**Figure 6.** (a)-(c) are the distribution of electric field $E_z$ along z-axis at three different moments $t/\xi = 1.72 \times 10^{-9}$ s, $t/\xi = 5.15 \times 10^{-9}$ s and $t/\xi = 1.37 \times 10^{-8}$ s, respectively. The red and black lines represent the electric field when scale factors are $\xi = 0.1$ and $\xi = 1$, respectively.

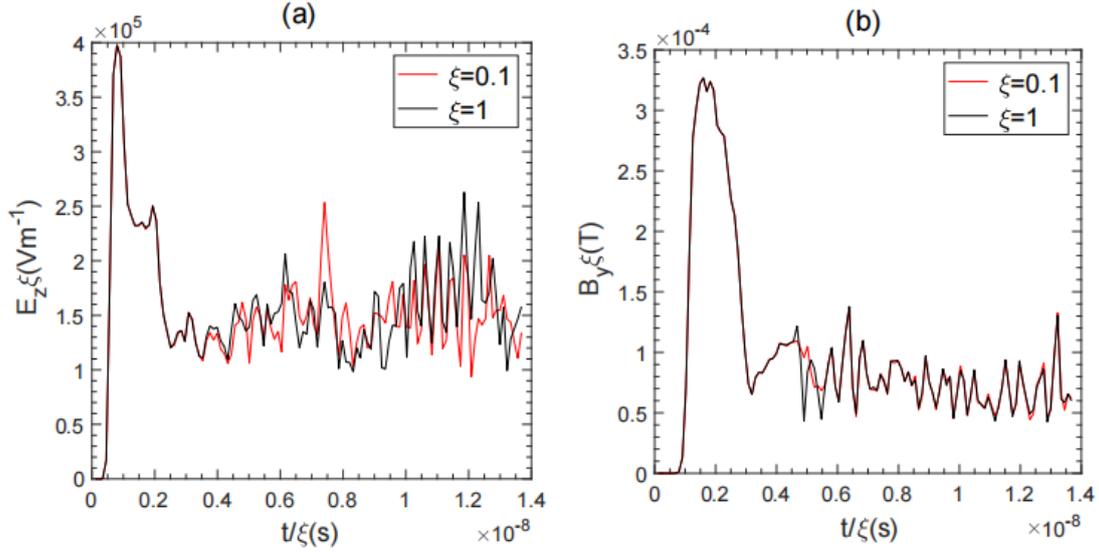

**Figure 7.** (a) and (b) show the evolution of the peak electric and magnetic fields in the simulation domain, respectively. The red and black lines represent the electric (magnetic) field when scale factors are $\xi = 0.1$ and $\xi = 1$, respectively.

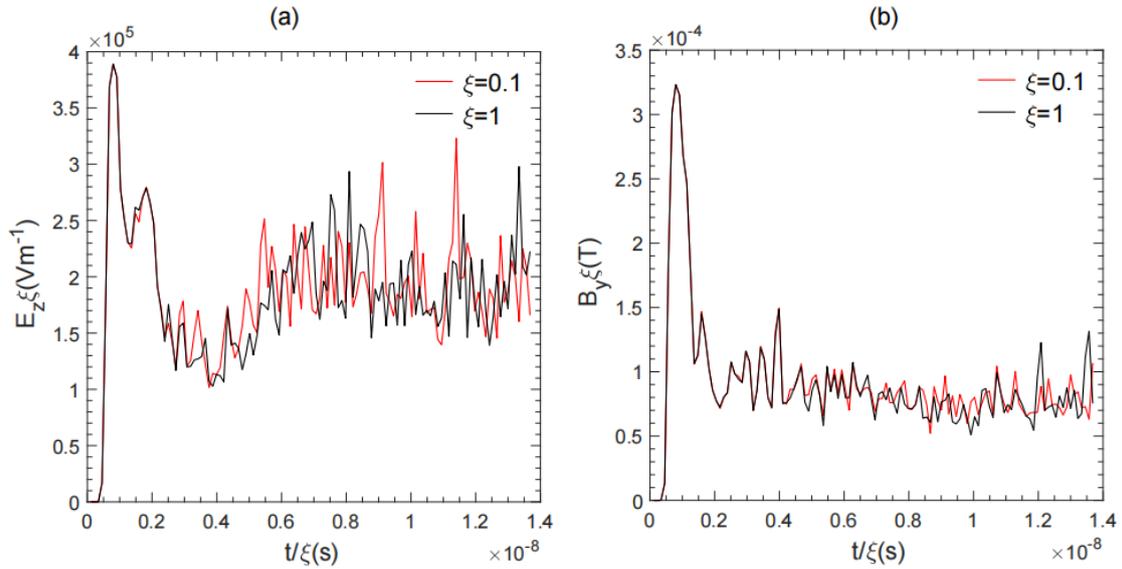

**Figure 8.** (a) and (b) show the evolution of the peak electric and magnetic fields in the simulation domain at higher pressure. The red and black lines represent the electric (magnetic) field when scale factors are $\xi = 0.1$ and $\xi = 1$, respectively.

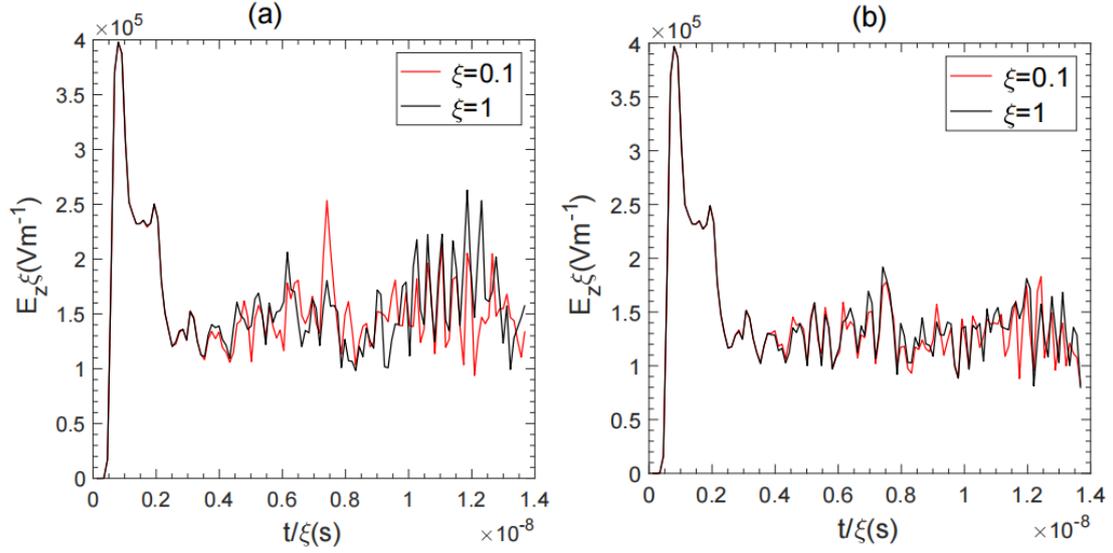

**Figure 9.** (a) the evolution of the evolution of the peak electric field for less macro particles in the simulation domain. (b) the evolution of the evolution of the peak electric field for more macro particles in the simulation domain. The red and black lines represent the electric (magnetic) field when scale factors are $\xi = 0.1$ and $\xi = 1$, respectively.

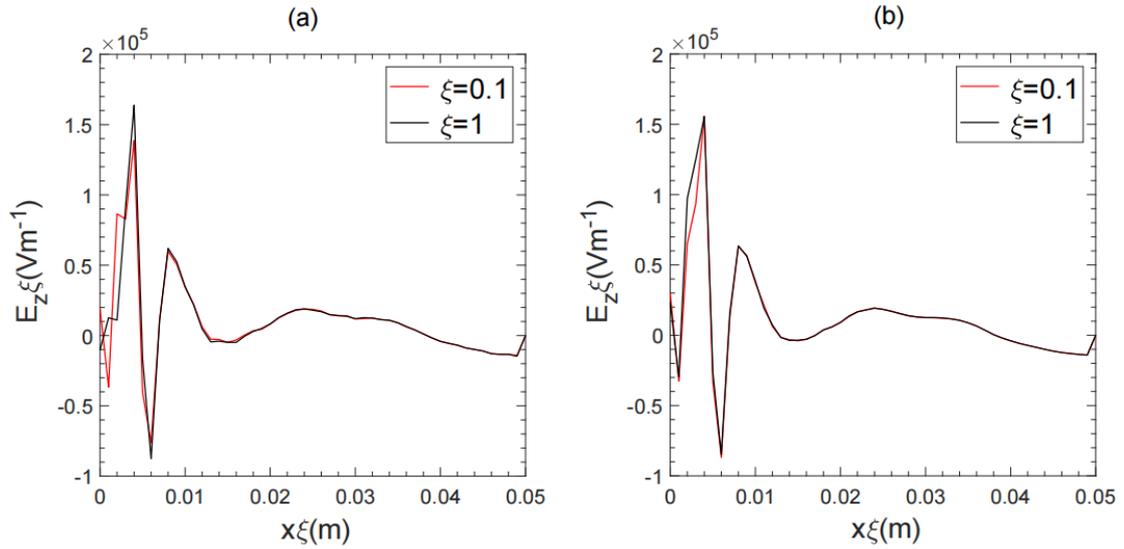

**Figure 10.** (a) is the distribution of electric field $E_z$ along z-axis at $t/\xi = 5.15 \times 10^{-9}$ s for smaller combination frequency. (b) is the distribution of electric field $E_z$ along z-axis at $t/\xi = 5.15 \times 10^{-9}$ s for larger combination frequency.. The red and black lines represent the electric field when scale factors are $\xi = 0.1$ and $\xi = 1$, respectively.

043502 (2005)

[15] A. Pukhov, S. Gordienko, S. Kiselev, and I. Kostyukov, Plasma Phys. Controlled Fusion https://doi.org/10.1088/0741-3335/46/12B/016 **46**, B179 (2004).

[16] S. Gordienko and A. Pukhov, Phys. Plasmas 12,043109 (2005)

[17] D. D. Ryutov and B. A.Remington, Plasma physics and controlled fusion, 48(3), 23 (2006)    doi:10.1088/0741-3335/48/3/l01

[18] T. C. Luce, C. C. Petty, and J. G. Cordey, Plasma Phys. Control. Fusion 50 (2008) 043001

[19] J. Chen, C. Zeng, J. Deng, and Z. Li, IEEE Trans. Nucl. Sci. **67**, 2353 (2020).

[20] J. Chen, J. Wang, Z. Chen, and Z. Ren, IEEE Trans. Nucl. Sci. **67**, 818 (2020).

[21] H. Huang and Y. Liu, IEEE Trans. Plasma Sci. **47**, 3631 (2019).

[22] A.J. Woods and T.N. Delmer, The Arbitrary Body of Revolution Code (ABORC) for SGEMP/IEMP (IRT CORP SAN DIEGO CA, 1976).

[23] R. Hockney and J. Eastwood, Computer Simulation Using Particles, New York:Taylor & Francis, Jan. 1988.

[24] C. K. Birdsall and A. B. Langdon, Plasma Physics via Computer Simulation (Series in Plasma Physics), New York:Taylor & Francis, Oct. 2004.

[25] A.J. Woods, W.E. Hobbs, and E.P. Wenaas, IEEE Trans. Nucl. Sci. **28**, 4467 (1981)

[26] H. T. Zhang, Q. H. Zhou, H. J. Zhou, Q. Sun, M. M. Sun, Y. Dong, W, Yang and J.S. Yao, Journal of applied physics (2021) DOI: 10.1063/5.0057841

[27] C. K. Birdsall. Charles, IEEE Transactions on plasma science 19.2, 65 (1991)

[28] Chanrion, Olivier, and Torsten Neubert., Journal of Computational Physics 227.15 (2008): 7222-7245.